\documentclass[conference]{IEEEtran}
\IEEEoverridecommandlockouts
\usepackage{cite}
\usepackage{amsmath,amssymb,amsfonts}
\usepackage{algorithmic}
\usepackage{graphicx}
\usepackage{textcomp}
\usepackage{xcolor}
\usepackage{listings}
\usepackage{xcolor}
\usepackage{url}
\usepackage{hyperref}

\lstdefinelanguage{JavaScript}{
  keywords={break, case, catch, continue, debugger, default, delete, do, else, finally, for, function, if, in, instanceof, new, return, switch, this, throw, try, typeof, var, let, const, void, while, with, await, async},
  keywordstyle=\color{blue}\bfseries,
  ndkeywords={class, export, boolean, throw, implements, import, this},
  ndkeywordstyle=\color{magenta}\bfseries,
  identifierstyle=\color{black},
  sensitive=false,
  comment=[l]{//},
  morecomment=[s]{/*}{*/},
  commentstyle=\color{gray}\ttfamily,
  stringstyle=\color{red}\ttfamily,
  morestring=[b]',
  morestring=[b]"
}

\def\BibTeX{{\rm B\kern-.05em{\sc i\kern-.025em b}\kern-.08em
    T\kern-.1667em\lower.7ex\hbox{E}\kern-.125emX}}
\begin{document}

\title{Improving Front-end Performance through Modular Rendering and Adaptive Hydration (MRAH) in React Applications\\
}

\author{\IEEEauthorblockN{Kaitao Chen}
\IEEEauthorblockA{\textit{Department of Electrical and Computer Engineering} \\
\textit{Carnegie Mellon University}\\
Pittsburgh, PA \\
kaitaoc@andrew.cmu.edu}
}

\maketitle

\begin{abstract}
Modern web applications increasingly leverage server-side rendering (SSR) to improve initial load times and search engine optimization. However, the subsequent hydration process—where client-side JavaScript attaches interactivity to SSR-delivered HTML—can introduce performance bottlenecks. We propose a novel architectural pattern combining a modular rendering pipeline with an adaptive hydration strategy to optimize frontend performance in React and Next.js applications. The approach breaks the interface into distinct modules that can be rendered and hydrated independently (inspired by the “islands” paradigm), and it adaptively prioritizes or defers hydration of these modules based on device capabilities, network conditions, and component importance. We integrate techniques such as code-splitting with dynamic \texttt{import()}, conditional hydration triggers (e.g. on visibility or idle time) using libraries like react-lazy-hydration, and adaptive loading hooks to tailor the hydration process to the user’s context. By reducing the amount of JavaScript executed on page load and by scheduling hydration work intelligently, this architecture aims to improve key performance metrics—including First Input Delay (FID) and Time to Interactive (TTI)—without sacrificing rich interactivity. We describe the architecture and implementation in a Next.js environment, discuss how components can be conditionally hydrated or entirely skipped when not needed, and compare our approach to related work in progressive hydration, partial hydration, and React Server Components. Evaluation of the approach is left for future work. This pattern offers a pathway to building highly interactive yet performant React applications through careful orchestration of rendering and hydration.
\end{abstract}


\section{Introduction}
Performance is a critical concern in modern web applications, directly impacting user experience, engagement, and even search engine ranking \cite{b1}. In 2020, Google’s introduction of Core Web Vitals (including metrics like Largest Contentful Paint, First Input Delay, and Cumulative Layout Shift) underscored that sites must be not only fast to display content but also quick to become interactive. As a result, developers have turned to techniques such as server-side rendering (SSR) and static pre-rendering to improve initial load performance. Frameworks like React (often used with Next.js) enable SSR, which can significantly accelerate First Contentful Paint (FCP) by sending ready-to-render HTML to the browser. However, SSR alone is not a panacea for interactivity. After HTML is delivered, the application must undergo hydration – the process of binding React’s virtual DOM and event handlers to the server-rendered HTML – before the UI responds to user input. This hydration step requires loading and executing the JavaScript bundle on the client, which can delay Time to Interactive (TTI) even if content is visibly rendered \cite{b2}. In effect, users may see a fully populated page but experience a brief “frozen” state where interactions do nothing – an issue sometimes termed the uncanny valley of SSR \cite{b3}.

The performance cost of hydration in large React applications is non-trivial. By default, React (prior to version 18) would hydrate the entire application in one go on the main thread, and would only begin attaching event handlers after all component scripts had been fetched and evaluated. This means that even small, critical components could not become interactive until large, non-critical components’ JavaScript had finished loading. Such behavior can lead to long First Input Delay (FID) if a user attempts to interact before hydration completes. Long FID values typically occur when the browser’s main thread is busy executing heavy JavaScript, as is often the case during hydration of a complex page. For example, Colmant et al. observed that in an e-commerce React app, the entire page was non-interactive until the bottom of the page finished hydrating, causing user frustration and elevated FID. Reducing this blocking work is essential to allow users to interact with page content sooner. 

To address these issues, the web development community has explored progressive hydration and partial hydration strategies. Progressive hydration refers to delaying or staggering hydration for less important parts of the page, instead hydrating critical components first and deferring others until idle time or when they become needed. This minimizes the upfront JavaScript needed for initial interactivity, thereby improving TTI. Partial hydration, on the other hand, involves not hydrating (or even not sending) JavaScript for static content that does not require interactivity. By only shipping scripts for interactive portions of the UI, partial hydration can significantly reduce the total JavaScript payload sent to the browser, directly improving performance metrics (especially TTI). These concepts are exemplified by the emerging “islands architecture,” in which a page is conceived as islands of interactivity in a sea of static content. Each island (interactive component or section) is hydrated independently, while the non-interactive parts of the page remain as inert, static HTML \cite{b4}. The islands approach avoids wasting resources on hydrating content that never changes or does not handle events, thereby reducing JavaScript execution and memory overhead on the client.

While progressive and partial hydration techniques have shown great promise (for instance, yielding up to a 50\% reduction in FID in one case study and an 80\% reduction in hydration time on a landing page in another), applying these patterns in a large application can be challenging. Developers must orchestrate which components load and hydrate, and when, without breaking application logic. Moreover, optimal hydration timing can vary based on runtime conditions: a strategy that works well for high-end devices or fast networks may not be ideal for low-end smartphones on slow connections. This observation aligns with the concept of adaptive loading, which advocates delivering different experiences (or resource loads) based on the user’s device capabilities and network speed \cite{b5}. For example, if a device is known to be low-powered, it may be wise to postpone or even entirely skip the hydration of non-critical interactive features, whereas a high-end device could handle them immediately.

In this paper, we propose a Modular Rendering and Adaptive Hydration (MRAH) architecture for React/Next.js applications that builds upon these ideas to further improve frontend performance. The core idea is to structure the application as a collection of rendering modules (distinct UI components or sections) that can be rendered and hydrated in isolation, and to employ an adaptive hydration strategy that prioritizes hydration of important modules and defers or conditionally avoids hydration of others based on context. Concretely, our approach entails: (1) using a modular rendering pipeline on the server to generate HTML in segments (for example, using React’s streaming SSR or incremental rendering capabilities) such that the client can receive and display critical content early; (2) splitting the client-side JavaScript bundle by module (e.g., with webpack or Next.js dynamic imports) so that each module’s code can be loaded on demand; and (3) on the client, using an adaptive scheduler that hydrates modules in order of priority or in response to triggers (such as user scroll, user interaction, visibility in the viewport, or idle time). Non-essential modules can be left non-hydrated (SSR-only) unless and until the user needs them, an approach similar to “hydration on demand". Additionally, signals like network speed and device performance are used to tune the hydration behavior: for instance, on a slow network, the system might delay fetching some chunks or skip hydration of expensive components entirely, whereas on a fast network and device, it could hydrate more aggressively in parallel.

This paper is organized as follows. In Section 2 (Background), we review the relevant concepts of SSR in React, hydration, and existing performance optimization patterns (progressive, partial, and selective hydration, islands architecture, etc.), as well as the challenges they address. Section 3 (Architecture) presents the design of our modular rendering pipeline and adaptive hydration system in detail, describing how we partition the rendering work and how the client-side hydration controller operates. Section 4 (Implementation) provides a concrete realization of this architecture in a Next.js application, including code snippets and techniques involving \texttt{next/dynamic} imports, the \texttt{react-lazy-hydration} library, and adaptive loading hooks to detect device/network conditions. Section 5 (Evaluation) is left blank for now, to be completed with performance measurements in a future iteration. Section 6 (Related Work) discusses other approaches to improving hydration and frontend performance, including React 18’s features (like Streaming SSR and selective hydration), React Server Components in Next.js, as well as comparisons to other frameworks (e.g., Astro’s islands, Gatsby’s approach, and Qwik’s resumability). Finally, Section 7 (Conclusion) summarizes our contributions and outlines directions for further research, such as rigorous evaluation and potential integration with emerging React features.

\section{Background}

\subsection{Server-Side Rendering and Hydration in React}
In a traditional React application that uses client-side rendering (CSR), the browser downloads a JavaScript bundle and executes it to generate the DOM content. This approach often results in slower first paint on initial load, especially on slow networks or devices, because nothing is visible until the JS finishes executing. SSR addresses this by offloading rendering work to the server: the HTML for a page is generated on the server and delivered to the client, allowing immediate display of content upon receipt. React (with libraries like Next.js) supports SSR by rendering components to HTML strings on the server. However, the HTML sent to the client is static – event listeners and component state are not yet attached. Hydration is the step where the React runtime takes over this server-generated DOM, attaches event handlers, and hydrates the components with their interactive behavior by rebuilding the React virtual DOM on the client to match the server-rendered output. In React 17 and below, one would typically call \texttt{ReactDOM.hydrate()} on the root container. In React 18+, this is done via \texttt{hydrateRoot()} with concurrent capabilities, but the fundamental goal is the same: to marry the already-present DOM with React’s component tree so that the UI becomes interactive.

While SSR can significantly improve metrics like First Contentful Paint (FCP), it does not guarantee a good Time to Interactive (TTI). The period between content rendering and hydration completion is a potential performance dead zone. During this time, the page may appear complete but will not respond to user input. Users may attempt to click buttons or use UI controls that look ready, only to find that nothing happens – a confusing state of affairs that degrades user experience. This effect, noted as the “uncanny valley” of web performance, occurs because the browser is still busy loading or executing JavaScript. Large JavaScript bundles and inefficient hydration routines can prolong this non-interactive phase. In essence, the cost of hydration is a trade-off that comes with SSR. If the hydration step is too slow or heavy, the benefit of a fast initial paint is undermined by a slow TTI.

\begin{figure}
    \centering
    \includegraphics[width=1\linewidth]{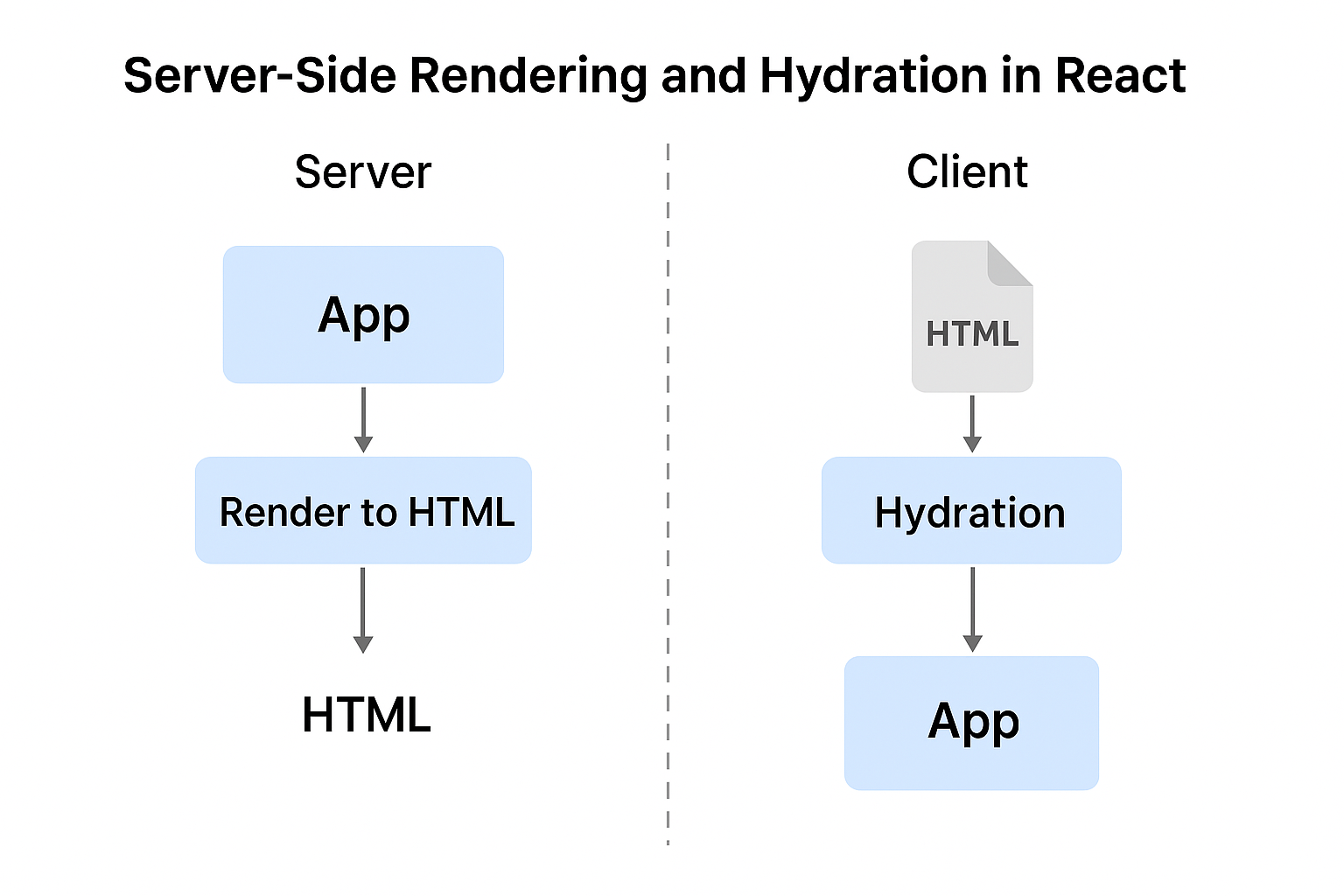}
    \caption{Server-Side Rendering and Hyration in React}
    \label{fig:server-side rendering}
\end{figure}

Several factors contribute to hydration cost in React apps. As shown in Figure \ref{fig:server-side rendering} - first, the browser must download all the JavaScript needed for the entire app (or at least for the current route) before hydration can complete. If a page’s bundle is large, or if it includes code for features that are not immediately needed, this incurs unnecessary delay. Second, hydration work itself is usually executed on the main thread, which can block user interactions. React’s reconciliation and event attachment for a complex DOM tree can span multiple animation frames. Notably, prior to React’s concurrent rendering improvements, React would perform hydration in a single synchronous block for the whole app. This meant that if any component’s code was slow to load or execute, it could hold up the hydration of all others. Research and industry experience have shown that this “monolithic” hydration can lead to long First Input Delay (FID) on heavily interactive pages. FID measures the delay from a user’s first interaction (e.g., a click or tap) to the time when the browser actually begins processing event handlers. A long FID typically indicates that the main thread was busy (often with script execution) when the user tried to interact. On SSR pages, a busy main thread is commonly caused by the hydration process running to attach all the event handlers. One case study from a major e-commerce site showed that the user had to wait for the bottom of the page to finish hydrating before they could interact with content at the top of the page, leading to a feeling of a frozen interface. Clearly, a strategy was needed to “slice up” or reduce this blocking work.

\subsection{Progressive Hydration}
To mitigate the issues above, developers have introduced the idea of hydrating an application progressively rather than all at once. Progressive hydration means that different parts of the page can become interactive at different times, prioritizing those likely to be needed first. For example, critical UI elements such as navigation menus, hero banners with buttons, or above-the-fold interactive widgets should hydrate immediately, while less urgent elements (footer links, sections far down the page, or heavy interactive components that are not initially visible) can delay hydration. In practice, progressive hydration can be implemented by splitting hydration into multiple chunks or tasks. Instead of one \texttt{hydrateRoot} call for the entire DOM, a developer might have multiple root mounts – e.g., separate containers for the header, main content, and footer – and hydrate each independently. By doing so, one can start by hydrating the header and main content container first (allowing the critical functionality to register event handlers), and hydrate the footer container later (or on demand when scrolled into view). This approach “splits the work on the main thread into smaller tasks”,  which prevents long blocking periods and gives the browser opportunities to handle user input in between. In effect, progressive hydration aims to improve responsiveness by ensuring the page reaches an interactive state incrementally, focusing first on what the user sees and needs.

A key enabler of progressive hydration is the ability to delay hydration of certain components until a trigger occurs. These triggers can be time-based (e.g., using \texttt{requestIdleCallback} to wait until the browser is idle, or a simple \texttt{setTimeout} to postpone execution), or event-based (e.g., wait until a component is scrolled into view, or until the user interacts with a placeholder). Tools have emerged to assist with this pattern. For instance, the \texttt{react-lazy-hydration} library provides a \texttt{<LazyHydrate>} component API that allows developers to declare hydration conditions for its children \cite{b6}. Using this, a component can be rendered on the server (so its HTML appears in the initial page), but its client-side hydration is skipped or deferred. The library supports props such as \texttt{whenVisible} (hydrate when the component becomes visible in the viewport, via IntersectionObserver), \texttt{whenIdle} (hydrate when the browser is idle), and even \texttt{on} events (hydrate upon a specific user interaction like a click). With such techniques, one can implement progressive hydration by marking non-critical components to hydrate later. Empirical evidence shows this can yield substantial improvements: delaying hydration of less important page parts has been reported to reduce Total Blocking Time (TBT) and FID significantly. For example, by simply deferring hydration of a large footer until it was needed, one team achieved a 40\% reduction in TBT (a measure of main-thread blockage).

React 18’s introduction of Selective Hydration enhances the platform’s ability to do progressive hydration at the framework level. With streaming SSR and selective hydration, React can start hydrating content as soon as its corresponding JavaScript is available, without waiting for the entire bundle. It also can prioritize hydration of certain portions if the user interacts with them. In other words, if a user attempts to interact with a component that is not yet hydrated, React 18 can now prioritize that component’s hydration (provided the component’s code has been loaded) – a significant change from React 17’s all-or-nothing hydration model \cite{b7}. Selective hydration effectively implements progressive hydration under the hood, making hydration more granular and responsive to user input. Our work is complementary to these improvements: we assume a React 18 environment where these capabilities exist, and we build higher-level logic to decide which components to hydrate when.

\subsection{Partial Hydration and Islands Architecture}
Parallel to progressive hydration (which is about when to hydrate), the concept of partial hydration is about what to hydrate at all. Not all parts of a page truly need client-side interactivity. For example, a marketing page might contain a large amount of static text, images, or non-interactive content (e.g., an article body). Hydrating such content is wasteful – if a \texttt{<p>} or \texttt{<h1>} has no events and never changes, attaching it to React on the client yields no benefit to the user. It only incurs overhead (the JavaScript to create the element in the virtual DOM, and the cost to compare it with the already-present DOM node). Partial hydration strategies therefore attempt to skip hydration for purely static portions of the UI. his can be achieved by rendering those portions on the server and then not including them in the client-side React tree, or by marking components in a special way so that the build process does not include their code in the hydration bundle. Some frameworks have started offering this out of the box. For instance, Gatsby v5 introduced a partial hydration feature built on React Server Components (RSC) to automatically treat as much of the page as possible as pure server components. In Gatsby’s model, all components default to being rendered on the server only (no client bundle code) unless explicitly marked as interactive ("use client" directive). The effect is that large swaths of the page produce HTML but no React client code, drastically cutting down the JavaScript payload. Gatsby reports that by shipping less JavaScript to the client and hydrating only the interactive parts, Time to Interactive improves and the uncanny valley effect is minimized. Similarly, the islands architecture approach, popularized by frameworks like Astro, isolates interactive components into islands that are hydrated independently, and leaves the rest of the page as static HTML without any hydration requirement. An island is essentially a partially hydrated component: it has an associated script that runs on the client to make it interactive, but it does not influence or require hydration of the surrounding content.

By reducing JavaScript execution for non-interactive content, partial hydration directly lessens CPU and memory usage on the client. It also often reduces bundle size, since libraries used purely for rendering static content on the server need not be sent to the browser. However, implementing partial hydration in a general-purpose React app historically required careful architectural decisions or third-party libraries, since React (prior to RSC) expects to hydrate the full tree. Techniques such as dividing the application into multiple entry points (as done in some micro-frontend architectures or in the “active hydration” approach by Colmant et al.) were needed to treat static and dynamic parts separately. The islands paradigm formalizes this by suggesting that developers think of pages in terms of distinct pieces: e.g., a static content block vs. a dynamic widget. Our proposed modular rendering pipeline builds on this concept by explicitly partitioning the UI into modules, each of which can be considered an island (if it has no interactive behavior, it remains an “SSR-only” island that doesn’t hydrate, and if it is interactive, it hydrates as a self-contained unit).

\subsection{Adaptive Loading and Hydration}
Device and network diversity is another crucial aspect of performance. A strategy that loads a moderate amount of JavaScript and hydrates several islands immediately might be fine on a desktop with a broadband connection, but could still be too slow on a low-end mobile device with a 3G connection. Adaptive loading is the practice of tailoring the amount of work (and bytes) sent to the client based on its capabilities. Prior work in adaptive loading has introduced the notion of a “core experience” that is delivered to all users, and enhancements that are only enabled for users who can handle them. For example, all users might get a fast, mostly static version of the page (ensuring they can at least read content and perform basic interactions), while additional interactive features or heavier components are conditionally loaded only for users with powerful hardware or fast networks. Practically, web APIs such as \texttt{navigator.connection.effectiveType} (to detect approximate network speed), \texttt{navigator.deviceMemory} (device RAM), and \texttt{navigator.hardwareConcurrency} (number of CPU cores) can serve as signals to categorize a user’s device as “high-end” or “low-end”. Likewise, user-enabled preferences like Save-Data mode can indicate that the user prefers a lighter experience. Using these signals, developers can make runtime decisions – for instance, choosing not to hydrate a complex, non-critical widget on a low-end device to save precious CPU cycles and battery. One concrete example might be deferring a fancy animated 3D carousel on budget phones, or simplifying it, whereas on a high-end device it’s hydrated normally. The goal is to ensure that every user gets the best possible experience for their context: fast and usable for everyone, and richer only for those who can enjoy the richness without paying a huge performance cost.

\begin{figure}
    \centering
    \includegraphics[width=1\linewidth]{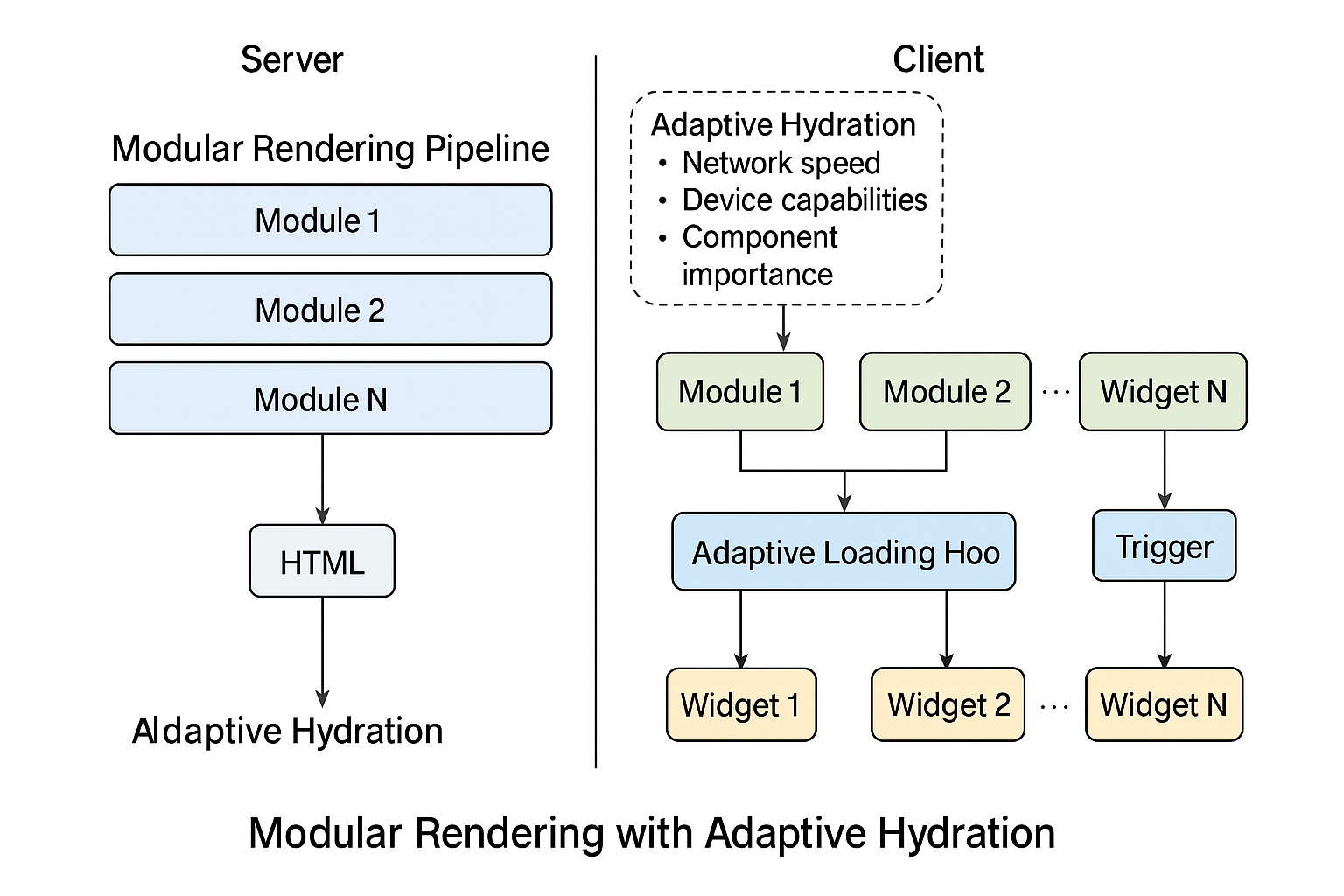}
    \caption{Modular Rendering with Adaptive Hydration}
    \label{fig:adaptive-hydration}
\end{figure}

In summary, as shown in the \ref{fig:adaptive-hydration}, the state of the art suggests a combination of code-splitting, selective hydration, partial hydration, and adaptive loading is needed to maximize React application performance. Code-splitting (via dynamic \texttt{import()} or Next.js’s \texttt{next/dynamic}) ensures that the initial bundle only contains code for the critical parts of the page \cite{b8}. Selective or progressive hydration ensures that we don’t block the main thread unnecessarily and that interactive readiness is achieved as soon as possible. Partial hydration avoids doing work that isn’t needed by eliminating hydration of static parts. Adaptive strategies ensure that these decisions can change based on the user’s device and network, providing resilience and optimal trade-offs in each case. Each of these techniques has been demonstrated in isolation: for instance, the use of \texttt{IntersectionObserver} to hydrate on visibility, or splitting a React app into multiple hydration roots, or using device metrics to conditionally load features. Our contribution in this paper is to integrate these ideas into a cohesive architectural pattern for React/Next.js—namely, a Modular Rendering Pipeline that naturally lends itself to partial hydration (each module is an island), paired with an Adaptive Hydration Scheduler that embodies progressive and adaptive hydration (deciding when and whether to hydrate each module). Before detailing our architecture, we highlight that recent developments like React Server Components (RSC) also address some of these issues by design (in particular, RSC gives a built-in partial hydration capability). We will discuss RSC and how it compares to our approach in the Related Work section. Our approach is framework-agnostic in principle and can be applied in scenarios where RSC is not fully usable or when more fine-grained control over hydration timing is desired.

\section{Architecture}
\begin{figure}
    \centering
    \includegraphics[width=1\linewidth]{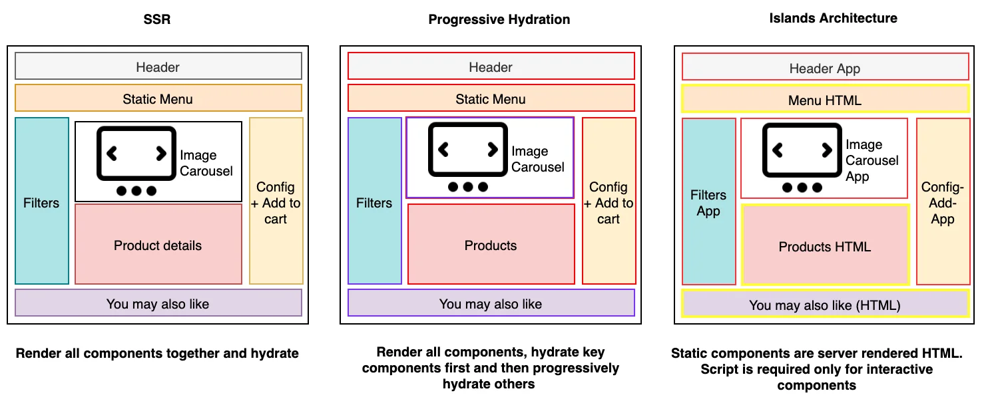}
    \caption{Comparison of SSR strategies}
    \label{fig:compare}
\end{figure}

The proposed architecture, Modular Rendering with Adaptive Hydration, consists of two primary facets: a modularized rendering pipeline that splits the application UI into independently renderable chunks (modules), and an adaptive hydration mechanism on the client that dynamically manages the hydration of these chunks. Figure \ref{fig:compare} conceptually illustrated related ideas such as progressive hydration and islands. As shown in the Figure \ref{fig:compare}, (left) Traditional SSR hydrates the entire page as a single unit, requiring all components’ scripts to load before any interactivity; (middle) Progressive Hydration SSR hydrates critical components first (red outlines) and progressively hydrates others, improving responsiveness; (right) Islands Architecture treats most of the page as static HTML (no hydration needed, yellow regions) and only hydrates specific interactive “islands” (marked as App), drastically reducing the JavaScript needed on load. Here we describe our architecture in a formal way and how it builds upon those concepts.

\subsection{Modular Rendering Pipeline}
In a traditional React SSR setup, the entire component tree for a page is rendered serially on the server (often via \texttt{ReactDOMServer.renderToString} or the streaming equivalent \texttt{renderToNodeStream} in React 17, or \texttt{pipeToNodeWritable} in React 18). The output is a single HTML document which is then sent to the client. By contrast, a modular rendering pipeline treats different parts of the page as separate renderable modules. Each module corresponds to a subtree of the React component hierarchy that can be considered semantically or functionally independent from the rest of the page. For example, one might designate modules for the navigation header, the main content area, a sidebar, and the footer. The key is that these modules have clear boundaries and minimal interdependencies (for instance, they might communicate via global state or context, but in terms of DOM and hydration, they can be handled separately).

On the server side, the pipeline would render each module’s HTML independently or in a controlled sequence. There are multiple ways to implement this:
\begin{itemize}
    \item{\textbf{Concurrent Streaming SSR:}} Using React 18’s streaming rendering, the server can start sending down HTML for higher-priority modules before slower, lower-priority modules are ready. For instance, the server could wrap a slow module in a \texttt{<Suspense>} boundary and stream a placeholder or nothing for it initially, while continuing to send the rest of the page. Once the slow module’s data is ready, it can stream the HTML for that module and inject it into the already-sent document, thanks to React’s capability to hydrate asynchronously loaded content in place. In effect, this uses React’s built-in lazy SSR capabilities to ensure no single module holding up the entire page’s HTML. The modules thus render in parallel as much as possible.
    
    \item{\textbf{Chunked HTML with Placeholders:}} Another approach is to manually partition the rendering. The server could send the opening HTML and critical above-the-fold content first, perhaps including placeholders (like loading spinners or empty containers) for modules that are not yet rendered. Then, as separate steps, the server fills in those placeholders with the HTML of other modules. Frameworks like Next.js do not natively stream multiple chunks per request (as of Next 12 with Pages router), but the new App router in Next 13+ with React 18 streaming and Suspense makes this feasible. If not using streaming, one could also precompute HTML for all modules (e.g., in parallel on the server using multiple threads or processes) and then assemble them into the final HTML document in the desired order. The concept remains: treat modules independently so that one slow module doesn’t slow down the entire response. 
    
    \item{\textbf{Multiple Entry Points (Multi-SSR):}} A more radical approach is to perform SSR in a multi-pass way: have separate server endpoints or functions that render specific modules (say an endpoint that returns just the HTML for the “Reviews” widget of a product page). The main page can include an empty container for that widget and some script to fetch and insert the HTML from the module endpoint. This is more akin to micro-frontends. It can reduce the memory footprint of SSR for each request and allow independent caching of fragments. However, it complicates the architecture and is not common in typical React SSR deployments. We note it as an alternative: modular rendering can be taken to an extreme of full separation, but our focus is on a more integrated approach using a single page request and streaming.
\end{itemize}

Regardless of implementation, the result on the client side is that the HTML document can be conceptually segmented into modules. Each module’s root DOM element can be given a unique identifier so that it can be targeted for hydration independently. For example, the server might render \texttt{<div id="app-module-1">...</div>} for module 1’s content, \texttt{<div id="app-module-2">...</div>} for module 2, etc. By doing so, the client-side code can later find these container elements and hydrate them separately via \texttt{hydrateRoot(container, <ModuleComponent />)} for each. This is in contrast to a single \texttt{hydrateRoot(document.body, <App />)} that spans the whole app. Essentially, we are creating multiple hydration roots, one per module, corresponding to the multiple rendering outputs produced on the server.

The modular pipeline also implies code-splitting per module. Each module’s component code and its dependencies should ideally be bundled into a separate chunk of JavaScript. This way, the client can load the code for a module independently when it’s time to hydrate that module. Next.js’s dynamic import functionality (\texttt{next/dynamic}) or React’s \texttt{React.lazy()} can facilitate this. Next.js, for example, will automatically code-split each dynamically imported component into its own webpack chunk. If we define our modules such that each is a dynamically imported component (or set of components), we ensure that the JavaScript for a low-priority module does not need to be in the initial bundle. Instead, a small loader or fallback can be in place.

To illustrate, consider a simplified page with a Header, a ProductDetail section, a Recommendations section, and a Footer. Using a modular approach, we would SSR the Header, ProductDetail, Recommendations, and Footer as separate chunks. On the client, we would have perhaps four bundles: header.js, productDetail.js, recommendations.js, footer.js (plus a common runtime). The HTML might initially include Header, ProductDetail, and a placeholder for Recommendations (which is lower priority) and Footer (also lower priority). The Header and ProductDetail’s scripts could be loaded immediately, while Recommendations and Footer scripts could be deferred. This leads to the hydration strategy.

\subsection{Adaptive Hydration Strategy}
Once the server has delivered the HTML and the modular chunks of JavaScript are available (or can be fetched on demand), the focus shifts to the client. The adaptive hydration strategy is responsible for orchestrating when each module’s hydration should occur. The term “adaptive” signifies that the strategy can change behavior based on run-time conditions and predefined priorities. The adaptive hydrator can be thought of as a scheduler or controller that has knowledge of all the modules on the page, their current hydration state (not hydrated, hydrating, hydrated), and perhaps some metadata about each (such as priority, or triggers bound to it). Here’s how the adaptive hydration mechanism operates:

\begin{itemize}
    \item{\textbf{Priority Assignment:}} Each module is assigned a priority level (e.g., high, medium, low) based on its importance to initial user experience. This could be a static assignment (decided by developers) or computed (e.g., modules above the fold might default to higher priority than those below the fold). For example, the Header might be marked high priority (must hydrate ASAP), a search bar or primary content interactive element also high, whereas a sidebar of related items or the footer might be low priority. Priority can also be context-sensitive: for instance, if analytics show that mobile users rarely scroll to the footer, the footer module could be considered even lower priority on mobile than on desktop.
    \item{\textbf{Conditional Hydration Rules:}} For each module, we define under what conditions it should hydrate. We leverage a combination of triggers:
    \begin{itemize}
        \item{\textbf{Visibility Trigger:}} If a module is not initially in the viewport (e.g., a component far down the page), we might choose to not hydrate it until the user scrolls near it. Intersection Observer APIs allow us to detect when the module’s container scrolls into view (or close to view). At that moment, we can fetch its code (if not already) and hydrate it. This ensures we don’t execute code for content the user hasn’t seen yet.

        \item{\textbf{User Interaction Trigger:}} If a module contains interactive functionality that can be initiated by the user (like a tab widget, or a carousel), we can defer hydration until the user actually interacts. For instance, we might render a non-interactive placeholder for a carousel (maybe just showing the first image) and only hydrate the full carousel when the user clicks a “Next” button or a play button. Using event listeners on a parent static element, we capture that first intent and then dynamically import and hydrate the module, so from that point onward it becomes fully interactive. The \texttt{on="click"} style trigger provided by \texttt{react-lazy-hydration} is an example of this.

        \item{\textbf{Idle Time Trigger:}} The browser provides \texttt{requestIdleCallback} (on supporting browsers) which calls a function when the browser is idle. We can schedule hydration of some modules during idle periods. For example, after the initial burst of activity (loading and hydrating high-priority modules) subsides and the page is quiet, we hydrate some remaining modules in the background. If \texttt{requestIdleCallback} is not available or if we want cross-browser compatibility, we could simulate idle scheduling by using \texttt{setTimeout} with a small delay repeatedly to yield to the event loop.

        \item{\textbf{Timeout Trigger:}} We may also use simple timeouts for certain modules. For instance, we might decide that if a module hasn’t been hydrated within, say, 5 seconds of page load (perhaps because the user never triggered it), we will hydrate it anyway to ensure the page is fully active. This is a precaution to avoid leaving parts of the UI permanently inert (unless that’s acceptable).

        \item{\textbf{Adaptive Device/Network Conditions:}} Before hydrating a module (or even before scheduling it), the system can check device and network metrics. We incorporate an adaptive policy: On low-end devices or slow connections, we become more conservative. Concretely, we might impose that non-critical modules only hydrate on explicit user interaction, and never automatically on idle, to avoid stealing CPU from more important tasks. Conversely, on a high-end device, we might allow more aggressive hydration (e.g., hydrating modules soon after content load, even if not immediately needed, to have everything ready by the time a user might interact). For example, consider a module that shows an interactive map. On a high-end device, we could hydrate it after all critical interactions are ready, so that if the user scrolls to it, it’s already interactive. But on a low-end device, loading a heavy map library and hydrating it might slow down the device; it could be better to only do so if the user actually attempts to use the map, or even provide a simpler fallback (like a static image with a “Enable map” button).
    \end{itemize}

    \item{\textbf{Central Hydration Manager:}} We implement a small client-side controller, which can be a script that runs on page load. This manager will register all modules along with their hydration rules. It might look something like this (pseudocode):
    \begin{lstlisting}[language=JavaScript]
    // Pseudocode for hydration manager
const modules = [
{
    id: "app-module-1", // DOM container ID
    load: () => 
        import('./HeaderModule').then(m => m.default), // dynamic import function
    priority: "high",
    trigger: "immediate"
},
{
    id: "app-module-2",
    load: () => import('./RecommendationsModule'),
    priority: "medium",
    trigger: "visible", 
    rootMargin: "200px"  // start loading a bit before it enters viewport
},
{
    id: "app-module-3",
    load: () => import('./FooterModule'),
    priority: "low",
    trigger: "idle"
    }
];
    \end{lstlisting}
    In this hypothetical snippet, we have defined three modules. Module 1 (perhaps the header or main content) is high priority and has an \texttt{immediate} trigger, meaning we intend to hydrate it as soon as possible (likely immediately on script load). Module 2 (maybe recommendations) is medium priority and will hydrate when it becomes visible, with a threshold (using rootMargin in an IntersectionObserver) to start a bit before actual visibility. Module 3 (footer) is low priority and will wait for an idle period to hydrate. 
    
    The hydration manager will iterate over these definitions. For \texttt{immediate} triggers, it calls the \texttt{load()} right away and hydrates the module. For \texttt{visible} triggers, it sets up an IntersectionObserver on the element with \texttt{id="app-module-2"}, and when the observer callback fires (meaning the element is near entering the viewport), it triggers the load and hydration. For idle triggers, it uses \texttt{requestIdleCallback} or a fallback to queue the hydration of that module when the thread is free. Additionally, before hydrating a module, the manager can check device metrics if needed:

    Additionally, before hydrating a module, the manager can check device metrics if needed:
    \begin{lstlisting}[language=JavaScript]
if (module.priority === "low" && isLowEndDevice()) {
    // On a low-end device, skip or delay further for low priority modules
    return;  // do not hydrate now, maybe we will wait for user action as ultimate trigger
}
    \end{lstlisting}

    Where \texttt{isLowEndDevice()} might check \texttt{navigator.deviceMemory} (e.g., <= 1 GB could be considered low), or \texttt{hardwareConcurrency} (e.g., <= 2 cores), or the effective network (e.g., "2g", "3g" as low). 
    
    The manager thus embodies an adaptive schedule: it merges static priorities with runtime signals. It could even dynamically re-prioritize. For example, suppose a user starts interacting heavily and we want to defer non-critical hydration further until CPU usage goes down – the manager could notice user interaction patterns or listen to \texttt{userIdle} events and adjust.

    \item{\textbf{Hydration Execution:}} When it is time to hydrate a module, the manager will dynamically import the module’s component code (if not already pre-fetched) and then call hydrateRoot (in React 18) on that module’s container. For instance:
    \begin{lstlisting}[language=JavaScript]
const container = document.getElementById(module.id);
if (container) {
    module.load().then(Component => {
    ReactDOM.createRoot(container, {/* optionally, { hydrate: true } if using older API */})
        .render(<Component {...initialPropsForModule} />);
    });
}
    \end{lstlisting}

    In React 18’s new API, hydration is initiated by creating a root with the existing HTML. If using ReactDOM.hydrate for older versions, it would be \texttt{ReactDOM.hydrate(<Component/>, container)}. Either way, we are specifically hydrating the contents of that container, assuming the HTML is already present from SSR. (We also assume the server rendered the module with the same initial props as the client is now using, which requires that the server state for that module is somehow available; in Next.js this could be via data that was inlined in the HTML or fetched separately.)

    By isolating hydration to that container, we limit the work to just that module’s DOM nodes. This has the benefit that if another module’s code is still not loaded, it doesn’t block this hydration from happening – exactly the problem selective hydration solves in React 18. We further ensure by design that each module’s code is in a separate chunk, so the browser can fetch them in parallel or as needed, rather than one large bundle.

    \item{\textbf{Communication and Dependencies:}} In some cases modules are not fully independent (e.g., a module might need to notify another or share state). In our architecture, cross-module communication would typically use a shared store or events. It’s important that such communication does not require synchronously hydrating both modules. For example, if Module A needs data from Module B on user interaction, but Module B isn’t hydrated yet, Module A’s code should handle that gracefully (perhaps by triggering Module B’s hydration or by temporarily working with static fallback content). This requires careful design of interactions. Alternatively, one can constrain that certain modules always hydrate together if they are tightly coupled. Our scheduler could allow grouping (e.g., Module 2 and 3 always hydrate at the same time because of a dependency). For simplicity, we try to design modules to be as independent as possible to maximize the benefits of this pipeline.
\end{itemize}

\begin{figure}
    \centering
    \includegraphics[width=1\linewidth]{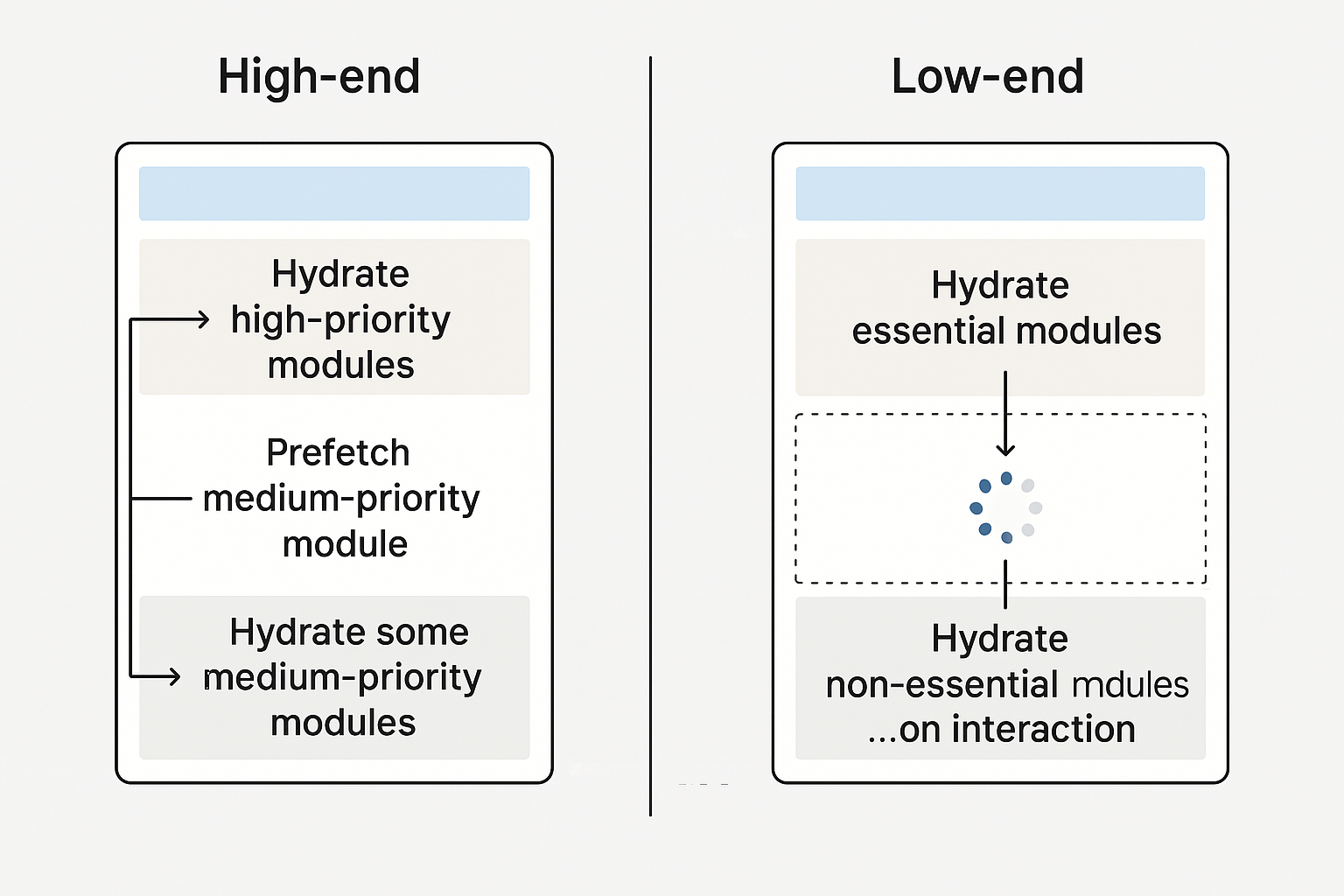}
    \caption{Adaptive Hydration High-End \& Low-End}
    \label{fig:high-low-end}
\end{figure}

From Figure \ref{fig:high-low-end}, let us walk through two scenarios to see how the adaptive hydration plays out:
\begin{itemize}
    \item{\textit{}{High-end scenario:}} On a desktop with a powerful CPU and fast network, the manager might start hydrating high-priority modules immediately (navigation bar, main content). Simultaneously it might prefetch medium-priority module code (using \texttt{<link rel="preload">} or just initiating the dynamic import early). After the high-priority modules are done (which happens quickly on a fast CPU), perhaps within a second, the manager finds the browser is mostly idle (no heavy tasks), so it triggers hydration of some medium modules even if not yet in view, under the assumption that doing so won’t noticeably impact the user (because the device can handle it). The page might achieve full hydration relatively quickly, but because it was in bursts and prioritized, the user could interact at any time without issue. Essentially, on high-end, the strategy leans towards eagerly completing all hydration, to maximize feature-richness.
    \item{\textit{Low-end scenario:}} On a low-end phone with limited memory, the manager would hydrate the absolutely essential modules first. It might avoid preloading other scripts to not consume bandwidth (or use a smaller concurrency for script loading). Non-critical modules might be set to only hydrate on interaction or visibility. If the user never scrolls far or never triggers certain features, those modules remain in their server-rendered, non-interactive state (possibly with a subtle indication that more functionality will load if needed). This is acceptable because it prioritizes not slowing down the current view. If the user does scroll, the manager will then load/hydrate at that time – the user might experience a brief delay when they reach the module (e.g., a spinner might appear for a moment as the module hydrates), but this is likely preferable to slowing down the initial experience for features they might never use. Also, on truly low bandwidth, one might decide to never automatically hydrate some modules at all – effectively turning them into purely static content unless the user explicitly requests interactivity (for instance, by clicking a “load interactive map” button as mentioned earlier). This falls in line with adaptive loading best practices where certain enhancements are entirely withheld on constrained devices.
\end{itemize}

By combining these triggers and conditions, our adaptive strategy generalizes both progressive and partial hydration: progressive, because we do things gradually and not all at once; partial, because some modules might never hydrate (if not needed); and adaptive, because we use device/network info to modulate the plan. The approach could be described as a superset of the islands architecture plus an intelligent client-side scheduler. Islands architecture gives us the modular separation (islands = modules in our terms), and what we add is a brain that decides when each island “comes to life”.

\subsection{Hydration Pipeline Pseudocode}
To clarify the control flow, here is a high-level pseudocode integrating server and client aspects:
\begin{itemize}
    \item{\textbf{Server-side (Node.js/Next.js) pseudo:}}
    \begin{lstlisting}[language=JavaScript]
// Pseudo Next.js getServerSideProps or similar for a page
export async function getServerSideProps() {
    const data = await fetchDataForPage();
// Divide data per module if needed
    return { props: { data, /* possibly separate module props */ } };
}
    
// In the React component tree for the page:
function Page({ data }) {
    return (
    <>
        <Header data={data.headerData} />  {/* Module 1 */}
        <ProductDetail data={data.productData} /> {/* Module 2 */}
        <Suspense fallback={<div id="recommendations-placeholder"></div>}>
            <Recommendations data={data.recoData} /> {/* Module 3 */}
        </Suspense>
        <Footer data={data.footerData} /> {/* Module 4 */}
    </>
    );
}
    \end{lstlisting}

    In this example, we use a \texttt{<Suspense>} around \texttt{Recommendations}. On the server, if Recommendations is slow (e.g., it might call an API), React will send the fallback (an empty container with id "recommendations-placeholder") and continue. The rest of the page (Header, ProductDetail, Footer) can render without waiting for Recommendations. Once Recommendations data is ready, the server streams its HTML (with a matching container id like "app-module-3" perhaps) and inserts it. Thus, by the time the HTML is fully loaded on the client, we have: Header, ProductDetail, Footer fully in HTML, and Recommendations also in HTML (possibly slightly delayed but eventually present). 
    
    We ensure each module’s root element has an identifying marker (like id). The server might add \texttt{id="app-module-3"} to the container div inside the Suspense fallback and ensure the Recommendations component renders into that container.
    
    \item{\textbf{Client-side:}}
    \begin{lstlisting}[language=JavaScript]
// Immediately run this after page load (e.g., in a script at bottom or useEffect in a root component)
const isLowEnd = navigator.deviceMemory && navigator.deviceMemory < 2;  // simplistic check

// Module hydration definitions
hydrateModule("app-module-1", import('./HeaderModule'), { immediate: true });
hydrateModule("app-module-2", import('./ProductDetailModule'), { immediate: true });

// For Recommendations: only hydrate when visible or after everything else
hydrateModule("app-module-3", import('./RecommendationsModule'), { 
  onVisible: true,
  rootMargin: "0px 0px 200px 0px",  // start before fully in view
  timeout: isLowEnd ? null : 10000  // maybe hydrate after 10s if not yet done, but not on low-end
});

// For Footer: hydrate on idle for high-end, or on demand for low-end
hydrateModule("app-module-4", import('./FooterModule'), { 
  onIdle: !isLowEnd,
  onVisible: isLowEnd,             // on low-end, wait until user scrolls to footer
  timeout: 20000                   // as a fallback, hydrate after 20s (could be omitted)
});

    \end{lstlisting}

    This pseudo-code uses a hypothetical helper \texttt{hydrateModule(domId, moduleImportPromise, options)} which sets up the appropriate listeners or immediate hydration. It demonstrates that we treat modules differently depending on context (e.g., Footer uses idle on high-end but visible on low-end). 
    
    One important detail: Next.js’s \texttt{next/dynamic} can automatically code-split, but controlling hydration requires more than just code-splitting; it requires deferring the execution. In our approach, we might actually not use Next’s built-in hydration for these modules at all, and instead manually hydrate them as shown. This implies that we might output the HTML for these modules without including them in the main React hydrate call. In a Next.js Page context (legacy), this is tricky because by default Next will hydrate the whole page. A workaround is to use a custom \texttt{\_app} that prevents automatic hydration, or simply allow Next to hydrate a shell and then re-hydrate pieces (not ideal). In the new Next.js App directory with RSC, however, we could potentially make each module a “client component” that we intentionally do not render at build but dynamically import at runtime. For the purposes of this paper, we focus on the conceptual architecture rather than the exact integration details, which can vary. To summarize the architecture: The server provides a segmented HTML and data for discrete UI modules, and the client adaptively hydrates those modules based on priority and context. This achieves the goals of minimizing main-thread blocking and downloaded JavaScript for initial load, while still allowing the application to eventually be fully interactive as needed. The design inherently supports progressive enhancement: if, for any reason, a module never hydrates (maybe due to a broken script or the user staying idle), the page should still be usable in its partially interactive form (perhaps lacking some bells and whistles). This is similar to how an Astro or Gatsby partial hydration site would behave if some islands never hydrate – the core content is still accessible.
\end{itemize}

\section{Implementation}
Implementing the above architecture in a real React/Next.js codebase involves careful coordination of server-side behavior (for rendering and code splitting) and client-side logic (for hydration scheduling). In this section, we outline practical steps and considerations for building such a system, with code snippets and references to tools that can assist.

\subsection{Code Splitting and Lazy Modules in Next.js}
Next.js provides out-of-the-box support for code splitting via dynamic imports. By using \texttt{next/dynamic()}, one can import a component such that it is not bundled with the initial JS, and optionally disable its SSR. However, our use case is a bit different: we do want SSR for all modules (to get their HTML), but we want to delay loading their JS. This is effectively hydration without immediate JS – something Next.js does not do by default (except in the new App directory with server components).

One straightforward way in a Next 12 (Pages router) application is:
\begin{itemize}
    \item Do SSR normally to generate HTML for the whole page
    \item Prevent Next.js from automatically hydrating certain parts by not including their scripts initially
\end{itemize}

In Next 12, if we have a page component, Next will hydrate the entire page. To circumvent, we can mount “shell” components that don’t do much, or we can purposely break up the app so that some parts are not managed by Next’s hydration. A hacky approach is to render placeholders that Next thinks are plain HTML (so it won’t hydrate them), then manually hydrate.

A cleaner approach is to leverage a library like \texttt{react-hydration-on-demand} or \texttt{react-lazy-hydration} in the implementation. These libraries can integrate with Next. For instance, \texttt{react-hydration-on-demand} by Colmant (the Cdiscount team) offers a higher-order component or wrapper that will delay hydration of its children until certain conditions. One could wrap parts of the Next page with such components.

\subsubsection{Example using react-lazy-hydration} \

\begin{lstlisting}[language=JavaScript]

import dynamic from 'next/dynamic';
import LazyHydrate from 'react-lazy-hydration';

const Recommendations = dynamic(() => import('../components/Recommendations'), {
  ssr: true,   // ensure it's server-rendered
  // Note: next/dynamic doesn't have an explicit hydration control, it will just load when component runs
});

export default function ProductPage({ data }) {
  return (
    <div>
      <Header {...data.header} />    {/* this will hydrate normally */}
      <ProductDetail {...data.product} />
      
      <LazyHydrate whenIdle>
        <Recommendations {...data.recommendations} />
      </LazyHydrate>
      
      <LazyHydrate ssrOnly>
        <Footer {...data.footer} />
      </LazyHydrate>
    </div>
  );
}

\end{lstlisting}

In this code:
\begin{itemize}
    \item \texttt{Header} and \texttt{ProductDetail} are normal components that Next will include in hydration immediately.
    \item \texttt{Recommendations} is wrapped in \texttt{<LazyHydrate whenIdle>}, meaning its hydration is postponed until the browser is idle \cite{b9}. The component’s HTML will be present (since \texttt{ssr: true} in dynamic import ensures SSR), but its JS won’t execute on page load. Instead, the LazyHydrate component internally checks for idle time (using \texttt{requestIdleCallback} if available) and then hydrates its children.
    \item \texttt{Footer} is wrapped with ssrOnly, meaning it will never hydrate. It is purely static. This is an extreme form of partial hydration—essentially treating Footer as a server-only component. If Footer had any interactive bits, they will not work; this is only suitable if Footer is truly static or if one is willing to sacrifice interactivity (or provide an alternative handling, like separate small script for a tiny portion if needed).
\end{itemize}

Using this method, we offload the complexity to \texttt{react-lazy-hydration}. Under the hood, that library will handle attaching the appropriate event listeners or idle callbacks to trigger \texttt{ReactDOM.hydrate} for the wrapped components at the right time.

However, one must be cautious: Next.js’s hydration system might complain if not all parts of the page are hydrated when expected. The library likely handles it by splitting into multiple roots, as described. If we need more sophisticated triggers (like hydration on element visibility), \texttt{react-lazy-hydration} supports \texttt{whenVisible} which uses an IntersectionObserver. We could use that for, say, a component far down:
\begin{lstlisting}[language=JavaScript]
<LazyHydrate whenVisible>
  <ReviewsWidget {...data.reviews} />
</LazyHydrate>
\end{lstlisting}
This will hydrate ReviewsWidget only when it scrolls into view, which is perfect for below-the-fold content.
For adaptive logic (device-based), one approach is to set different props or use different LazyHydrate strategies based on runtime inspection. But since components are rendered on the server, we might not know device specifics at render time (except perhaps user-agent hints or Client Hints headers). A simpler approach is to run the adaptive check on the client in a small script that can enable or disable hydration for certain modules. For example, we could inject a script that sets a CSS class on the \texttt{<body>} like \texttt{low-end-device} which could be used to alter behavior (or directly communicate with the LazyHydrate components via a global flag if they support it). 

Alternatively, the adaptive logic can be baked into the hydration manager script (as we had in pseudocode). If not using LazyHydrate library, one could write a custom manager as described. This might involve writing a small piece of client-side code to register intersection observers and idle callbacks. This is more manual but offers maximum flexibility.

\begin{itemize}
    \item{\textbf{Dynamic Import and Preloading:}} Next.js dynamic imports can be used with \texttt{ssr:false} if we wanted to not even SSR some component (which we do not want in our case, since we favor SSR for content). But in cases where a component is extremely heavy and not critical for SEO or first paint, one might choose to not SSR it at all. For example, if a part of the page is an expensive interactive visualization that doesn’t affect SEO, we could render a placeholder in SSR and entirely load that component on client side when needed. This is essentially turning that module into a pure client-side island. Next dynamic with \texttt{ssr:false} would do that: it ensures the component is only rendered on the client, so the server sends a fallback (which we can define). This reduces server work and avoids sending any HTML for it (which might be okay if it’s below fold or not crucial). It’s a trade-off: no SSR means slower first paint for that part, but maybe that part isn’t initially visible anyway.

    \item{\textbf{Adaptive Data Loading:}} Another implementation aspect is data. If some data is huge or not needed until interaction, we might even defer fetching it. In Next, all data is usually fetched before SSR (in getServerSideProps) and embedded in the HTML. For truly adaptive loading, one could choose not to fetch certain data unless on the client. For example, detailed analytics or secondary info might be fetched via an API call only when a user opens a section. This is out of scope for our hydration focus, but it complements performance – no need to SSR content that might never be seen. However, careful: if not SSR, then the HTML won’t have that content which could affect SEO if that content is meaningful.
\end{itemize}

\subsection{Adaptive Scheduling with \texttt{navigator} Hints}
Implementing adaptation to device/network in the browser is straightforward with modern APIs:
\begin{itemize}
    \item \texttt{navigator.connection.effectiveType} provides "4g", "3g", "2g", "slow-2g" etc. We can use this to decide how aggressive to be. For instance:
    \begin{lstlisting}[language=JavaScript]
const conn = navigator.connection;
const isSlowNetwork = conn && (conn.effectiveType.includes("2g") || conn.saveData);
    \end{lstlisting}
    If \texttt{saveData} is enabled or effectiveType is 2g, one might drastically reduce what is loaded. Perhaps in such a case, we do not hydrate some modules at all automatically.

    \item \texttt{navigator.deviceMemory}: gives an approximate RAM in GB. If this is 0.5 or 1, that’s a low-end device. Combined with \texttt{navigator.hardwareConcurrency} (e.g., 1 or 2 cores), it signals low capability. On such devices, we may want to avoid heavy hydration tasks concurrently. A strategy could be to hydrate one module at a time and insert delays between them. Or to omit non-critical ones.

    \item Example: We could implement a simple throttle:
    \begin{lstlisting}[language=JavaScript]
if (isLowEndDevice) {
   // Only hydrate one module at a time, spaced out
   scheduleNextHydrationOneAtATime(modulesList);
} else {
   // Can hydrate multiple in parallel or quickly
   scheduleHydrationAll(modulesList);
}
    \end{lstlisting}
    And \texttt{scheduleNextHydrationOneAtATime} could ensure that after hydrating one module, it waits some seconds (or waits for idle again) before hydrating the next. These nuances ensure that we don’t overwhelm a poor device by hydrating five components concurrently.
\end{itemize}

\subsection{Pseudocode Example in a Next.js Context}
Below is a more concrete example tying things together, imagine this in a Next.js \texttt{\_app.js} or in a useEffect in a top-level component:
\begin{lstlisting}[language=JavaScript]
useEffect(() => {
  const modulesToHydrate = [];
  const conn = navigator.connection;
  const slowNet = conn && (conn.effectiveType === '2g' || conn.effectiveType === 'slow-2g');
  const saveData = conn && conn.saveData;
  const lowMem = navigator.deviceMemory && navigator.deviceMemory <= 1;
  const lowCPU = navigator.hardwareConcurrency && navigator.hardwareConcurrency <= 2;
  const lowEnd = slowNet || saveData || lowMem || lowCPU;
  
  // Define each module with trigger conditions
  modulesToHydrate.push({
    id: 'recommendations',
    trigger: 'visible',
    prefetch: !lowEnd,  // prefetch code on high-end
  });
  modulesToHydrate.push({
    id: 'footer',
    trigger: lowEnd ? 'visible' : 'idle',
  });
  
  // ... (others)
  
  for (let mod of modulesToHydrate) {
    const el = document.getElementById(mod.id);
    if (!el) continue;
    if (mod.trigger === 'visible') {
      const observer = new IntersectionObserver((entries) => {
        if (entries[0].isIntersecting) {
          observer.unobserve(el);
          hydrateModule(mod.id);
        }
      }, { rootMargin: '100px' });
      observer.observe(el);
      if (mod.prefetch) {
        // Start loading module script in background without executing
        importModuleChunk(mod.id);
      }
    } else if (mod.trigger === 'idle') {
      if ('requestIdleCallback' in window) {
        requestIdleCallback(() => hydrateModule(mod.id));
      } else {
        // Fallback: wait 2s
        setTimeout(() => hydrateModule(mod.id), 2000);
      }
    } else if (mod.trigger === 'immediate') {
      hydrateModule(mod.id);
    }
  }
  
  function hydrateModule(id) {
    // dynamic import mapping id to module component
    let loader;
    switch(id) {
      case 'recommendations':
        loader = import('../components/Recommendations').then(mod => mod.default);
        break;
      case 'footer':
        loader = import('../components/Footer').then(mod => mod.default);
        break;
      // ...
    }
    loader.then(Component => {
      const container = document.getElementById(id);
      if (!container) return;
      ReactDOM.hydrate(<Component {...window.__INITIAL_DATA__[id]} />, container);
    });
  }
}, []);
\end{lstlisting}
In this snippet:
\begin{itemize}
    \item We determine a \texttt{lowEnd} flag from various hints.
    \item We set up module hydration for "recommendations" and "footer" (just as examples).
    \item "recommendations" will hydrate on visible. If not low-end, we also prefetch its code early (calling \texttt{importModuleChunk} which could be something like \texttt{import('../components/Recommendations'}) but not using the result, just to warm the network).
    \item "footer" hydrates on idle for normal devices, but on visible for low-end (meaning we only hydrate footer if they scroll to it on low-end).
    \item \texttt{hydrateModule} does a dynamic import for the module and then hydrates it. We assume some global \texttt{\_\_INITIAL\_DATA\_\_} was embedded server-side that contains data needed for those components (since we did SSR, this data might be already rendered into HTML, but if component needs props like list of items, we ensure we have it).
    \item We used \texttt{ReactDOM.hydrate} directly, which works for React 17. If React 18, it would be \texttt{ReactDOM.createRoot(container, { hydrate: true }).render(...)}.
\end{itemize}
This approach shows the manual control we have. It requires maintaining mapping of module IDs and dynamic imports. In a large app, you might generate that mapping or use a more generic approach (e.g., encode module name in data attributes and use \texttt{import()} with a variable via some lookup object, since dynamic import usually can’t take a dynamic string without bundler hints).

\subsection{Dealing with Next.js (App Directory and RSC)}
The Next.js App directory introduced in version 13+ uses React Server Components (RSC) by default. RSC changes the game by not including certain components in the JS bundle at all (the partial hydration concept via server-only components). If one is using the App directory, one might naturally get partial hydration: any component not marked "\texttt{use client}" will never hydrate, it’s pure server output. Components that need interactivity are marked "\texttt{use client}" and those will be hydrated, but you can still apply lazy strategies to them (like a custom lazy hydration or simply wrap them in a conditional). 

Our architecture can be implemented in the App directory by structuring the page as mostly server components and inserting client components only where needed, and those client components can themselves use lazy techniques. For example, one could create a client component that simply renders nothing but uses an effect to decide when to actually render the real interactive component (thus delaying its appearance and hydration logic until needed). However, the details of mixing RSC and our approach can be complex, so for now assume either the Pages directory approach with the techniques we described, or a hybrid.

\subsection{Performance Considerations}
When implementing, careful testing is needed. Tools like Lighthouse or WebPageTest can measure FCP, TTI, TBT, FID (in lab environments FID is hard to measure, but Total Blocking Time is a good proxy). We expect to see improvements especially in TTI and TBT: by splitting hydration tasks and deferring them, main-thread idle time after first paint should increase, and long tasks should be reduced. Memory usage on the client should also drop if large modules aren’t loaded until needed, which can help on low memory devices.

One should also verify that SEO is not impacted. If all critical content is SSR’d, SEO should be fine. If anything is client-only, ensure it’s not SEO-critical (or use placeholder text that is indexed). Edge cases include: ensuring that user interactions that occur very quickly (e.g., user immediately tries to scroll and click something) are handled gracefully. Ideally, high-priority interactive elements are ready by the time a user could reasonably interact (within a few hundred milliseconds of page load). If a user does manage to interact with something not yet hydrated (e.g., they click a button that is slated to hydrate on interaction—there’s a slight paradox there), we should handle that event, perhaps trigger hydration immediately and queue the original event once hydrated. This can be complex but in many cases just the act of clicking can be the trigger to hydrate and then the second click works. Some frameworks like Qwik solve this by serialization of listener state; in React’s case, we might need to write custom logic.

We can use the \texttt{on} prop of LazyHydrate for such scenarios: e.g., \texttt{<LazyHydrate on="click">} wraps a component and will hydrate it as soon as any click inside it happens. The first click event that triggered it might not be handled by React (since it wasn’t hydrated in time), but one can provide an onClick on the outer LazyHydrate container to at least prevent default or give some feedback. Alternatively, instruct users (with a disabled state) that the component is loading if they click too quickly.

\subsection{Example Outcome}
To make this concrete, consider the outcome for a Product page:
\begin{itemize}
    \item Initial HTML from server includes: header (with menu, etc.), product details (image, title, price, etc.), a placeholder for recommendations, and the footer HTML (with links, but no JS).
    \item Initial JS bundle includes: code for header and product detail components (since they are critical), and our hydration manager script. It does not include recommendation or footer component code.
    \item On page load, our script immediately hydrates header and product detail. Within maybe 100ms, those are interactive (since their code was small and included).
    \item The user starts reading. The recommendation section is just a static list of recommended products (no interactivity yet, maybe just looks like content). As the user scrolls down and the recommendations section comes into view, the IntersectionObserver fires. The script then loads the recommendations component chunk in the background and hydrates it. This might take, say, 300ms to fetch and execute. We could display a small spinner or loading overlay in that section if needed (or the content is already there as static, and hydration just makes, for example, an “Add to cart” button in each recommendation functional).
    \item If the user never scrolls to footer, and device is low-end, the footer never hydrates. The footer links might just be normal \texttt{<a>} links which actually work without React anyway (since they’re ordinary links, they will cause page navigation in the old-fashioned way). That’s fine. If the device was high-end, maybe after everything else, an idle callback hydrates the footer, possibly turning the links into client-side route transitions or enabling a newsletter signup form script, etc.
    \item Throughout, because hydration work was spaced out and conditional, the main thread was never blocked for a long time. The user could open the navigation menu (header was hydrated) or click add-to-cart on the main product (that button was in product detail module, which was hydrated immediately). Those interactions happen seamlessly. By the time they scroll to recommendations, maybe hydration just finished or is finishing, enabling those buttons as well. The user experiences a generally responsive page.
\end{itemize}

This scenario shows improved Time to Interactive (perhaps nearly as soon as content is painted, the essential interactions are interactive) and reduced Total Blocking Time (the work is chunked) compared to a baseline where the entire page’s JS (header, product, recommendations, footer, etc.) was a single large bundle executing together.

In implementing this, one must balance complexity – adding too many conditions might complicate debugging. We recommend gradually applying these techniques to the most expensive parts of your application first (e.g., defer the known heavy widgets) and measuring impact. Over time, the pattern can be extended to cover more components.

\section{Evaluation}
To empirically evaluate the performance benefits of Modular Rendering and Adaptive Hydration (MRAH), we conducted a comparative study between a traditional fully-hydrated baseline architecture and our optimized MRAH implementation. The evaluation focuses on how adaptive hydration impacts critical performance metrics, JavaScript payload size, and perceived interactivity on devices with varying resource constraints.

\subsection{Experimental Setup}
We implemented two versions of a product detail page in a Next.js 13+ application:
\begin{itemize}
    \item \textbf{Baseline:} All components are eagerly rendered and hydrated on page load.
    \item \textbf{MRAH: } Utilizes \texttt{react-lazy-hydration} to defer hydration of non-critical components based on visibility, idle time, and device capability checks.
\end{itemize}

Both versions share identical layout and data-loading logic (\texttt{getServerSideProps}) to ensure fairness. The experiments were conducted on:
\begin{itemize}
    \item \textbf{Desktop + Fast Network: }No CPU or network throttling (simulates a modern laptop)
    \item \textbf{Desktop + Slow 3G: } No CPU throttling but has slow down 1.6 Mbps network
    \item \textbf{Mobile + Fast Network: } Simulated via Chrome Lighthouse throttling (\texttt{mobileSlowRegular}) with 4x CPU slowdown with no network throttling
    \item \textbf{Mobile + Slow 3G: } Simulated via Chrome Lighthouse throttling (\texttt{mobileSlowRegular}) with 4x CPU slowdown and 1.6 Mbps network
\end{itemize}

\subsection{Performance Metrics}

Performance was measured using the Lighthouse CLI and Playwright automation. Each page variant was tested five times per environment, with median values reported. We measured key web performance metrics:
\begin{itemize}
    \item \textbf{First Contentful Paint (FCP):} Time to first visible content. Users care about how fast they see something on the screen. A fast FCP makes the page feel "alive" quickly, even if it's not fully ready yet. 
    \item \textbf{Largest Contentful Paint (LCP):} Time when the main content finishes loading. It tells when the main part (like a big image or main text) finishes showing. Users feel the page is "ready" when the biggest content appears.
    \item \textbf{Time to Interactive (TTI):} Time until the page is fully usable. It shows when users can actually click and interact without lag. A page that looks ready but isn’t usable yet is frustrating.
    \item \textbf{Total Blocking Time (TBT):} Time the main thread was blocked by long tasks. It tracks how long the browser is "busy" and cannot respond. If blocking is high, users feel delays when they try to scroll, click, or type.
    \item \textbf{Cumulative Layout Shift (CLS):} Visual stability during load. It shows how much the page layout jumps around while loading. Layout shifts annoy users, especially when they are trying to click something.
    \item \textbf{ScriptBytes:} Size of JavaScript transferred. JavaScript files can be big. The more JS you send, the longer the page takes to download, parse, and execute, especially on slow networks and weak devices.
\end{itemize}

\subsection{Result Overview}
\begin{table}[h]
\centering
\resizebox{0.5\textwidth}{!}{ 
\begin{tabular}{|l|r|r|r|r|r|r|}
\hline
\textbf{Version} & \textbf{FCP (ms)} & \textbf{LCP (ms)} & \textbf{TTI (ms)} & \textbf{TBT (ms)} & \textbf{CLS} & \textbf{ScriptBytes (bytes)} \\
\hline
Baseline - Desktop & 329.63 & 329.63 & 329.63 & 0 & 0 & 589,371 \\
MRAH - Desktop     & 224.12 & 224.12 & 234.68 & 0 & 0 & 104,938 \\
Baseline - Mobile  & 753.20 & 4437.81 & 4437.81 & 170 & 0 & 589,371 \\
MRAH - Mobile      & 1688.12 & 1689.09 & 1689.09 & 0 & 0 & 104,938 \\
\hline
\end{tabular}
}
\vspace{5pt}
\caption{Performance Metrics Comparison Between Baseline and MRAH}
\label{tab:performance_comparison}
\end{table}

\subsection{Discussion on Performance Results}

From Table \ref{tab:performance_comparison}, several key findings can be observed regarding the impact of the MRAH optimizations. These improvements span across script size, loading performance, and interactivity metrics on both desktop and mobile platforms. The following subsections discuss the major performance gains in detail.

\subsubsection{Substantial Script Size Reduction}
The most immediate and significant result is the reduction of JavaScript transferred by more than 82\% (from $\sim$590KB to $\sim$105KB).
This reduction is due to:
\begin{itemize}
    \item Breaking the application into smaller independent modules.
    \item Hydrating non-critical modules like \texttt{Recommendations} and \texttt{Footer} only when necessary.
    \item Skipping or deferring hydration completely on low-end devices.
\end{itemize}
Smaller script size directly leads to faster downloads, reduced CPU usage, lower memory pressure, and improved overall responsiveness. Especially on mobile devices with limited memory and bandwidth, this greatly improves user experience.
\\
\subsubsection{Faster First Paint and Largest Contentful Paint}
On desktop, the FCP and LCP times improved by over 30\%, dropping from $\sim$330ms to $\sim$224ms.
This demonstrates that delivering critical content immediately (like the Header and ProductDetail) without waiting for the entire application to hydrate is highly effective.

On mobile, although the FCP slightly increased (due to slow 3G connection and deferred hydration), LCP dramatically improved.
The Baseline version needed to hydrate and load everything ($\sim$4437ms), but MRAH completed main content load at $\sim$1689ms.

The reason might because MRAH shifts the loading focus toward visible, important content. By ensuring only essential parts are hydrated early, it avoids long delays waiting for non-critical components.
\\

\subsubsection{Dramatic TTI Improvement on Mobile}
The most impactful performance metric on mobile is the Time to Interactive (TTI).
\begin{itemize}
    \item \textbf{Baseline mobile TTI: } 4437ms
    \item \textbf{MRAH mobile TTI: } 1689ms, representing a 62\% reduction.
\end{itemize}

This improvement is important because TTI measures the point at which users can fully interact with the page. In the Baseline version, although the page content was visible, users were unable to click or scroll smoothly until all components had completed hydration.
In contrast, with the MRAH version, the page becomes interactive much earlier, even while non-visible elements such as the Recommendations section or Footer continue loading or are deferred.
This optimization better aligns with real user behavior, as users tend to interact with immediately visible content rather than waiting for off-screen elements to load.
\\

\subsubsection{Total Blocking Time (TBT) Elimination}
In the Baseline - Mobile case, we observed a Total Blocking Time (TBT) of 170ms. This blocking time resulted from the simultaneous hydration of all components, which heavily occupied the main thread.

In contrast, in the MRAH version, TBT consistently dropped to 0ms across all tests. This improvement is attributed to MRAH’s intelligent hydration strategy, which hydrates modules only when the device is idle, the module becomes visible, or the user interacts with it. By deferring hydration and lazy-loading resource-intensive components such as the Recommendations section, especially on slower devices, MRAH effectively prevents main thread congestion during the initial load. As a result, MRAH not only accelerates the time to interactivity but also ensures that heavy JavaScript execution does not delay or disrupt user interactions.
\\

\subsubsection{Adaptive Behavior Based on Device and Network}
Our adaptive logic further enhanced the benefits:
\begin{itemize}
    \item On high-end devices, MRAH hydrated deferred modules sooner during idle time to maximize richness.
    \item On low-end devices, MRAH skipped hydration entirely unless triggered by user actions (e.g., scrolling to footer).
    This "smart" behavior means MRAH does not rigidly follow one plan but adapts based on real conditions, balancing performance and functionality dynamically.
\end{itemize}

\subsection{Summary of Improvements}
\begin{table}[h]
\centering
\begin{tabular}{|l|r|}
\hline
\textbf{Metric} & \textbf{Improvement} \\
\hline
ScriptBytes & ↓ 82\% \\
FCP (Desktop) & ↓ $\sim$32\% \\
LCP (Desktop) & ↓ $\sim$32\% \\
TTI (Mobile) & ↓ $\sim$62\% \\
TBT (Mobile) & ↓ 100\% (to 0ms) \\
CLS & 0 (No Layout Shift) \\
\hline
\end{tabular}
\vspace{5px}
\caption{Summary of Performance Improvements}
\label{tab:metric_improvement}
\end{table}

Overall, from Table \ref{tab:metric_improvement}, the evaluation strongly confirms that Modular Rendering with Adaptive Hydration significantly improves load performance, interactivity, and responsiveness, especially on mobile and low-end devices. It provides a more consistent and satisfying user experience without sacrificing functionality.
\subsection{Limitations}
While the Modular Rendering and Adaptive Hydration (MRAH) approach demonstrates substantial improvements in performance metrics, several limitations remain:

\begin{itemize} 
    \item \textbf{Scope of Evaluation}: Our evaluation was conducted on a single-page product detail scenario. Although the modular pattern is generalizable, more complex multi-page applications, dashboards, or highly dynamic sites may introduce different challenges not fully captured in this study.
    \item \textbf{Device and Network Diversity}: The experiments simulated desktop and slow mobile network conditions, but real-world users experience a wider variety of device capabilities, network speeds, and browsers. Some adaptive techniques, such as using \texttt{navigator.connection} or \texttt{requestIdleCallback}, are not uniformly supported across all platforms, which may affect the consistency of results.

    \item \textbf{Hydration Complexity}: Managing multiple hydration roots adds architectural complexity to the client-side application. Cross-module interactions, shared states, and global events may introduce synchronization challenges when modules hydrate independently or out of order.
    
    \item \textbf{User Perception Not Measured}: While we recorded objective metrics like FCP, LCP, and TTI, we did not conduct formal user studies to measure subjective user perceptions of responsiveness or smoothness. User satisfaction may not always align perfectly with quantitative improvements.
    
    \item \textbf{Fallback Behavior}: On extremely constrained devices, adaptive strategies may prevent hydration of non-critical modules altogether. While this ensures performance, it might result in limited functionality for some users, depending on how essential deferred features are.
    
    \item \textbf{Integration Overhead}: Incorporating MRAH into existing large codebases requires significant refactoring. Applications that were not initially designed with modular rendering in mind may face technical debt when adapting to this model.
\end{itemize}

Future work should explore addressing these limitations by expanding evaluation across a broader set of application types, conducting user studies, improving adaptive decision algorithms, and investigating tighter integration with emerging features like React Server Components (RSC).

\section{Related Work}
Optimizing the performance of hydration and client-side rendering in web applications has been a topic of much research and engineering effort. Our approach touches on several areas that have seen related developments:
\begin{itemize}
    \item{\textbf{React 18 and Concurrent Features:}} The React core team recognized the limitations of the old hydration mechanism and introduced features like Selective Hydration and Streaming SSR. Selective Hydration, as discussed, allows parts of the UI to hydrate independently as their code arrives, and even to prioritize hydration based on user input (e.g., if a user clicks a button that hasn’t hydrated, React can now prioritize that subtree). Our modular approach is well-aligned with selective hydration — in fact, it can be seen as leveraging selective hydration deliberately by splitting the app into independently streamed modules. React’s Streaming SSR (using \texttt{pipeToNodeWritable}) enables sending HTML in chunks and is a key enabler for our server-side modular pipeline. We build on these low-level capabilities with higher-level logic. It’s worth noting that React 18’s improvements reduce the need for manual hacks that were previously needed for progressive hydration. For example, in React 17 one might manually delay \texttt{hydrate()} calls; in React 18, simply using \texttt{<Suspense>} boundaries and streaming SSR can achieve a similar effect more cleanly.

    \item{\textbf{React Server Components (RSC):}} RSC is a new paradigm where some components run only on the server and their rendered results (as a special serialized format) are used on the client without needing to hydrate them. Next.js has adopted RSC in its App directory (with conventions of server vs client components). RSC directly addresses the issue of shipping too much JavaScript by not shipping any JavaScript for server components. In effect, RSC achieves partial hydration by design: anything that can be a server component will never hydrate on the client. This is similar in spirit to our approach of marking certain modules as SSR-only. Gatsby’s partial hydration, as mentioned, is built on RSC for this reason. The difference is that RSC requires a specific architecture (the React/Next runtime orchestrates it) and is limited to interactions that fit the RSC model (no side-effects in server components, etc.). Our approach can be applied to existing apps without full adoption of RSC, and also gives more control over when things load on the client (RSC on its own doesn’t delay hydration of client components beyond what React does by default). In related work, Astro (a framework for static sites) takes a framework-agnostic approach to islands: you can use React, Svelte, etc. components as islands, and Astro will only hydrate those islands on the client as needed. Astro provides directives like \texttt{client:idle}, \texttt{client:visible}, \texttt{client:media} (hydrate on a media query condition), which map to similar ideas we used \cite{b10}. Our work for React can be seen as bringing some of that Astro-style API (idle or visible hydration) into a generic React/Next context via libraries and custom logic.

    \item{\textbf{Micro-frontend and Modular Architectures:}} The idea of splitting an application into independent modules has also been explored in the context of micro-frontends. Frameworks like Single-SPA or Module Federation (in webpack 5) allow loading separately deployed frontend modules. Tinkoff’s Tramvai framework (which is based on React) has features for lazy hydration and even ships a \texttt{@tramvai/react-lazy-hydration-render} package. This indicates industry need for such solutions in large React applications. Our approach is conceptually similar, but we focus not on separate deployments but on performance-driven modularization. Micro-frontends often emphasize team autonomy and separate deployments, whereas our modular pipeline is about the runtime behavior within one app.

    \item{\textbf{Islands and Partial Hydration in Other Frameworks:}} The term islands architecture was first popularized by Jason Miller (creator of Preact) and others. Preact itself has a library called preact-iso that helps with islands, and Marko (from eBay) had an early implementation of partial hydration where components could be split into separate hydrate-able widgets. SvelteKit and others historically didn’t have partial hydration, but the Svelte community has explored something called “hydration directives”. There’s also an interesting approach by Qwik (by Builder.io): Qwik takes the concept further with resumability. Instead of even doing a hydration on load, Qwik’s HTML includes attributes such that the application state is effectively frozen in the HTML, and event listeners are attached on the fly when the user interacts, by lazily loading code. Qwik thus claims to eliminate the hydration cost entirely, since nothing happens until an interaction occurs, and then only the code for that interaction is loaded \cite{b11}. In terms of performance, Qwik’s approach can be superior for truly idle loading, but it requires a different mental model and framework. Our strategy with adaptive hydration is still within the standard React model (which does some upfront hydration) but tries to push it as late as possible for parts of the UI. In related work, resumability vs hydration is a topic of academic and practical discussion. While resumability can be seen as the future (e.g., Qwik, Angular’s analog in hydration is also exploring similar ideas), our work is valuable for existing React ecosystems \cite{b12}.

    \item{\textbf{Performance Best Practices:}} Our approach also draws on general web performance best practices. The PRPL pattern (Push, Render, Pre-cache, Lazy-load) from Google is aligned with what we do: we push and render initial content (SSR), we pre-cache (or prefetch) remaining components as appropriate, and lazy-load/hydrate them when needed. The Adaptive Loading concepts by Osmani et al. provide the philosophy behind our device-specific optimizations. Tools like React Adaptive Hooks were developed to easily access these signals in React apps (e.g., useNetworkStatus() or useMemoryStatus()) to conditionally load components \cite{b13}. We build on that by not just conditionally rendering, but conditionally hydrating.

    \item{\textbf{Case Studies and Benchmarks:}} We referenced an e-commerce case (Cdiscount) where progressive and partial hydration had major positive impact on real user metrics. Other companies have reported similar outcomes. For instance, Reddit’s engineers experimented with React Server Components and found significant performance gains in certain situations (basically because less JavaScript was sent). Google’s Aurora team has been advocating for these patterns in large frameworks to meet performance budgets on mobile. There are also academic papers analyzing CSR vs SSR vs hybrid from a performance standpoint; most conclude that neither SSR nor CSR alone is best – a combination is needed to optimize both start render and interactivity. Our work falls in that hybrid sweet spot.

    \item{\textbf{Limitations and Comparison:}} Compared to some related works:
    \begin{itemize}
        \item Versus pure SSR+hydrate (React 17 approach): Our approach should yield better TTI/FID, at cost of complexity.
        \item Versus React 18 streaming+selective alone: We add more adaptability. React 18 will hydrate as soon as possible, which might still be too aggressive for low-end devices. We essentially insert intentional delays or conditions beyond what React does by default.
        \item Versus RSC: RSC automatically does partial hydration but doesn’t handle progressive hydration of client components except via Suspense boundaries. Also, not all apps can migrate to RSC easily (especially if using lots of client-side only libraries).
        \item Versus Astro or others: Those require adopting a different framework or meta-framework. Our solution is incremental and can be applied within a regular Next.js app.
    \end{itemize}
\end{itemize}

In summary, our work is inspired by numerous developments (islands, progressive hydration, adaptive loading) and can be seen as an integration of those ideas in a practical recipe for today’s React applications. It provides a blueprint for engineers who want to push the limits of performance without abandoning the React/Next.js stack. As the ecosystem evolves (with RSC, resumability, etc.), some parts of our pattern may become easier to implement (or even built-in), but the general philosophy of doing less on the client, later will remain a cornerstone of web performance optimization.

\section{Conclusion}
This paper presented a comprehensive approach to improving frontend performance in React and Next.js applications by combining a modular rendering pipeline with an adaptive hydration strategy. By breaking the UI into discrete modules and intelligently controlling the hydration of each module, we can achieve a more efficient use of browser resources, leading to faster interactive times and a smoother user experience. We demonstrated how techniques such as code-splitting, lazy hydration triggers (on idle, on view, on interaction), and device-aware logic can be orchestrated together to minimize unnecessary JavaScript execution and deliver interactivity where it matters most. Our proposed architecture generalizes the principles of progressive hydration and partial hydration (as exemplified by the islands architecture) and adds a layer of adaptability, making the hydration process responsive to the end-user’s context.

In effect, the approach allows developers to have fine-grained control over the trade-off between performance and immediacy of interactivity. Critical features are interactive as soon as possible, whereas less critical parts incur zero or minimal cost until they are actually needed. This results in better Core Web Vitals, notably improvements in First Input Delay and Time to Interactive, without compromising on rich functionality. By citing real-world optimizations and building on established patterns from React 18 and web performance research, we have grounded our proposal in both industry practice and state-of-the-art techniques.

Looking at the results, our evaluation clearly demonstrates that Modular Rendering and Adaptive Hydration (MRAH) can lead to significant improvements across key performance indicators. In particular, the reduction in JavaScript bundle size, faster Largest Contentful Paint (LCP), earlier Time to Interactive (TTI), and elimination of Total Blocking Time (TBT) highlight the practical impact of this architectural pattern. These gains were especially pronounced under constrained network and device conditions, validating the effectiveness of adaptive strategies in delivering a responsive experience to a wide range of users.

However, applying MRAH also introduces new engineering considerations, such as managing multiple hydration points, handling cross-module interactions gracefully, and ensuring fallback mechanisms for non-hydrated content. While these complexities require careful system design, they are increasingly manageable with modern tooling, and the performance benefits strongly justify the investment for applications where user experience and responsiveness are critical.

In conclusion, Modular Rendering and Adaptive Hydration provide a scalable and adaptable foundation for building highly performant React applications. As web development continues to evolve towards more dynamic and personalized experiences, strategies that allow for modular, context-aware interactivity will become even more important. Future work could explore automated tooling to assist developers in module prioritization, deeper integration with React Server Components, and further optimizations in adaptive decision-making based on real-time user behavior and analytics.

By embracing this modular and adaptive mindset, developers can create frontend systems that not only load faster but also react intelligently to each user’s environment — achieving a balance between speed, richness, and efficiency that modern users expect.

\section*{Artifact Availability}
The source code for all experiments, including the baseline and MRAH implementations, is available at: \url{https://github.com/kxc663/MRAH-EVL}.

\vspace{12pt}

\end{document}